\documentstyle[twocolumn,seceq]{jpsj}
\def\v#1{\mib #1}
\def\sbox#1{\mbox{\scriptsize #1}}
\def\Hca{H_{\sbox{c0}}}
\def\Hcb{H_{\sbox{c1}}}
\def\Hcc{H_{\sbox{c2}}}
\def\Hs{H_{\sbox{s}}}

\def\runtitle{Mganetization Plateaus in 1-d $\v{S}=1/2$ Heisenberg Model with Dimerization and Quadrumerization}
\def\runauthor{Wei {\sc Chen}, Kazuo {\sc Hida} and Hiroki {\sc Nakano}}
\title
{
Magnetization Plateaus in One Dimensional $\v{S}=1/2$ Heisenberg Model with Dimerization and Quadrumerization
}

\author
{
Wei {\sc Chen}\footnote{E-mail: chenwei@riron.ged.saitama-u.ac.jp}, Kazuo {\sc Hida} and Hiroki {\sc Nakano}$^1$
}

\inst
{
Department of Physics, Saitama University, Urawa, Saitama 338-8570\\
$^1$ Institute for Solid State Physics, University of Tokyo, Roppongi, Tokyo 106-8666}

\recdate
{
\today
}

\abst
{
The one dimensional $S=1/2$ Heisenberg model with dimerization ($1-j$) and quadrumerization ($\delta$) in the magnetic field is studied by means of the numerical exact diagonalization of finite size systems and the conformal field theory. It is found that the magnetization plateau at half of the saturation value exists for $\delta \neq 0$. For $\delta = 0$, this model is described by the conformal field theory with central charge $c=1$ at this value of magnetization. The critical exponent $\nu$ which characterizes the $\delta$-dependence of the width of the plateau is calculated using the level spectroscopy method. The $j$-dependence of the critical exponent $\nu$ is found to be non-monotonic and discontinuous at $j = 0$. The effective theory of the magnetization plateau is also presented for various limiting cases.
}

\kword
{
Heisenberg chain, spin gap, exact diagonalization, magnetization plateau, conformal field theory, central charge, scaling dimension, dimer gas model.
}

\begin{document}
\sloppy
\maketitle

\section{Introduction}
Recently, the Heisenberg chains with a variety of spatial structures have attracted great interest. Although the uniform $S=1/2$ antiferromagnetic Heisenberg chain can be solved exactly \cite{bh} and its ground state is known as a gapless spin liquid, various spatial structures such as spin-Peierls dimerization, zig-zag chain or ladder structure can drive this ground state into spin gap states.

One the other hand, the magnetization plateaus in one dimensional Heisenberg chains have been also studied by a lot of researchers\cite{kh,ts2,tone,mo,to2,to,naru,ca1,ca2,fled1,fled2,na,AK,WS} quite recently. This state can be regarded as the field induced spin gap state in which spins are partly quenched by the magnetic field and remaining spins form a state with a finite energy gap {\it i.e.} the magnetization plateau. It is evident that such magnetization plateaus can be realized in various non-homogeneous Heisenberg chains.

In the previous work\cite{we}, two of the present authors (Chen and Hida)  investigated the ground state phase diagram of the $S=1/2$ Heisenberg model with coexisting dimerization and quadrumerization. Similar polymerized Heisenberg chains are also studied by many authors\cite{ca2,fled1,fled2,we,zvi1,zvi2,zvi3,tak}. In this paper, we further investigate the magnetization plateaus in this model using the numerical exact diagonalization method and the conformal field theory.

This paper is organized as follows. In the next section, the model Hamiltonian is defined. The numerical results for the magnetization plateaus are presented in {\S}3. Analyzing the exact diagonalization data by the conformal field theory, we calculate the central charge $c$ and the critical exponent $\nu$ characterizing the opening of the plateau. The effective theory in several limiting cases is also pesented. The last section is devoted to summary and discussion.

\section{Model Hamiltonian}
The Hamiltonian of the one dimensional dimerized and quadrumerized $S=1/2$ Heisenberg chain in the magnetic field is given by
\begin{eqnarray}
{\cal H} &=&  j\sum_{l=1}^{2N} \v{S}_{2l-1}\v{S}_{2l} + \sum_{l=1}^{2N}(1+(-1)^{l-1} \delta)\v{S}_{2l}\v{S}_{2l+1} \nonumber \\
& -& g\mu_{\rm B}H\sum_{l=1}^{4N}S_{l}^{z},
\end{eqnarray}
where  $1-j(-\infty \le j \le \infty)$ and $\delta (-1 \le \delta \le 1)$ represent the degree of dimerization and quadrumerization, respectively.  The magnetic field, the electronic $g$-factor and the Bohr magneton are denoted by $H, g$ and $\mu_{\rm B}$, respectively. In the following, we take the unit $g\mu_{\rm B}=1$ and $\delta \ge 0$ without loss of generality and assume the periodic boundary condition($\v{S}_{4N+1}=\v{S}_{1}$). For $j > 0$, this model can be regarded as the quadrumerized antiferromagnetic-antiferromagnetic alternating chain while for $j < 0$, it can be regarded as the quadrumerized ferromagnetic-antiferromagnetic alternating chain. This model approaches the dimerized spin-1 antiferromagnetic Heisenberg chain in the limit $j \rightarrow -\infty$.
According to the criterion by Oshikawa et al.\cite{mo}, it is possible that the present system exhibits the magnetization plateau at  $m^{z}=1/4$ where $m^{z}$ is the magnetization per site defined by $m^{z}=\frac{1}{4N}\displaystyle\sum_{l=1}^{4N}S_{l}^{z}$.

\section{Numerical Results}
\subsection{Magnetization Plateaus}

We use the exact diagonalization method to calculate the magnetization curve of the present model. Typically the magnetization curve has the form shown schematically in Fig. \ref{fig1} with a plateau at $m^{z}=1/4$. The $\delta$-dependence of the 4 critical fields $\Hca, \Hcb, \Hcc$ and $\Hs$ are shown in Fig. \ref{fig2} for (a)$j=5$, (b) 0.5, (c) $-0.5$ and (d) $-5$. The critical field $\Hca$ corresponds to the energy gap in the absence of magnetic field, $\Hcb$ to the energy gap between the lowest energy with $M^{z}=N$ and that with $M^{z}=N-1$, $\Hcc$ the energy gap between the lowest energy with $M^{z}=N+1$ and that with $M^{z}=N$ and $\Hs$ is the saturation field.  Here $M^{z}$ is the total magnetization defined by $M^{z}=4N m^z$. The magnetization plateaus at $m^{z}=1/4$ open between the $\Hcb$ and $\Hcc$ as shown in the Fig. \ref{fig2}(a-d) by the symbols $\diamond$ and $\circ$, respectively. The values of $\Hcb$ and $\Hcc$ in Fig. \ref{fig2} are obtained by the Shanks' \cite{shanks} extrapolation of the results of the Lanczos exact diagonalization for $4N=12,16,20,24$ and 28. The points at which $\Hca$ vanishes are determined using the method of Ref. \citen{we}. The symbol $\bullet$ represents $\Hca$ obtained similarly from the data for  $4N=8,12,16,20$ and 24. The thick solid line is the saturation field $\Hs$ which can be calculated analytically as follows.

The saturation field is given by the energy difference $\Delta E$ between the lowest energy with $M^{z}=2N-1$ and that with $M^{z}=2N$\cite{kh,hi}. The latter is simply given by the energy of the fully polarized state. The former is the plane wave state of a single inverted spin. The energy difference  $\Delta E$ is given by the lowest eigenvalue of the matrix,
\begin{equation}
\left[
\begin{array}{cccc}
-\frac{j}{2}-\frac{1-\delta}{2}&\frac{j}{2}&0&\frac{1-\delta}{2} \mbox{e}^{{\rm -i}4k} \\
\\
\frac{j}{2}&-\frac{j}{2}-\frac{1+\delta}{2}&\frac{1+\delta}{2} & 0 \\
\\
0&\frac{1+\delta}{2}&-\frac{j}{2}-\frac{1+\delta}{2}&\frac{j}{2}\\
\\
\frac{1-\delta}{2} \mbox {e}^{{\rm i}4k}&0&\frac{j}{2}&-\frac{j}{2}-\frac{1-\delta}{2} \\
\end{array}
\right] .
\end {equation}
Where the wave number is denoted by $k$. We find that $\Delta E$ is the lowest for $k=0$. Thus, the saturation field $\Hs$ is obtained as,
\begin{equation}
\Hs=-\triangle E=\frac{(j+2)+\sqrt{j^{2}+4\delta^{2}}}{2}.
\end{equation}

In view of these figures, it is quite likely that the plateau at $m^{z}=1/4$ opens for infinitesimal $\delta$. Therefore we expect that $\delta=0$ is the critical point. It should be also noted that the similar plateau is found in the $S=1$ dimerized antiferromagnetic Heisenberg chain not only theoretically\cite{tone,to2} but also experimentally\cite{naru}. This can be regarded as the $j \rightarrow -\infty$ limit of the present model. Comparing Fig. \ref{fig2}(a,b) and (c,d), the width of the magnetization plateaus glows more rapidly for $j >0$ than for $j < 0$. This suggests that the critical exponent $\nu$ of the plateau width defined by $\Hcc-\Hcb \sim \delta^{\nu}$ behaves differently depending on the sign of $j$. In the next subsection, we study this point in more detail by  the level spectroscopy\cite{nomura} analysis of the Hamiltonian with $\delta=0$ based on the conformal field theory.

\subsection{Critical Exponent $\nu$ of the Magnetization Plateau}

Assuming that the ground state with $m^{z}=1/4$ is critical and conformally invariant at $\delta=0$ , we estimate the central charge $c$ from the  Lanczos exact diagonalization data. It is known that the finite size correction to the ground state energy is related with the central charge as follows\cite{ca,HW,IA},
\begin{equation}
\frac{1}{2N}E_{\sbox{g}}(N,M^{z}) \cong \varepsilon (m^{z})-\frac{\pi}{6}cv_{\sbox{s}}\frac{1}{(2N)^{2}},
\label{ke}
\end{equation}
where  $v_{\sbox{s}}$ is the spin wave velocity, $E_{\sbox{g}}(N,M^{z})$ is the ground state energy with system size $4N$ and magnetization $M^{z}$ and $\varepsilon (m^{z})$ is the ground state energy per unit cell in the thermodynamic limit with corresponding value of $m^z$. It should be noted that the size of the unit cell is 2 for $\delta=0$.
In order to determine $v_{\sbox{s}}$, we also calculate the excitation energy with fixed  wave vector $k$ under the periodic boundary condition with $M^{z}=N$. The system sizes are $4N=12,16,20,24$ and 28. We denote the lowest energy with fixed $k$ by $E_{k}(N,M^{z})$ and the lowest one among them is the ground state energy $E_{\sbox{g}}(N,M^{z})$. For $j > 0$, the ground state has $k=0$ irrespective of the value of $M^{z}$. For $j < 0$, on the other hand, the ground state has $k=0$ for even $M^{z}$ and $k=\pi$ for odd $M^{z}$\cite{ts1}.

Thus, the velocity $v_{\sbox{s}}$ is estimated by
\begin{equation}
v_{\sbox{s}}=\frac{2N}{2\pi}[E_{k_{1}}(N,M^{z})-E_{\sbox{g}}(N,M^{z})],
\end{equation}
where $k_{1}$ is equal to $2\pi/(2N)$ for $j > 0$ and $j < 0$ with even $M^{z}$, whereas it is $\pi-2\pi/(2N)$ for $j < 0$ with odd $M^{z}$.

From Eq. (\ref{ke}), the gradient $A$ of the plot $E_{\sbox{g}}(N,M^{z})/2N$ versus $1/(2N)^2$ is equal to $\pi cv_{s}/6$ which has negligible size dependence. The finite size central charge $c$ is determined as $6A/(\pi v_{\sbox{s}})$. The extrapolation procedure for $c$ is shown in Fig. \ref{fig3} which yields $c \simeq 1.019\pm0.001$ for $j=0.5$. Figure \ref{fig4} shows the $j$-dependence of the central charge $c$ which indicates that $c$ is close to unity everywhere. The error bars due to the extrapolation procedure are less than the size of the symbols. Thus we may safely assume that the critical line $\delta=0$ with $m^z=1/4$ is described by the Gaussian model with conformal charge $c=1$ given by,
\begin{equation}
{\cal H_{\sbox{G}}}= \frac{1}{2\pi} \int {dx [v_{s}K (\pi\Pi)^{2}+\frac{v_{s}}{K}(\frac{\partial \phi}{\partial x})^{2} ]},
\end{equation}
where $\phi$ is the boson operator compactified as $0 \leq \phi < \sqrt{2}\pi$ and $\Pi$ is the momentum density conjugate to $\phi$ which satisfies $[\phi(x),\Pi(x')]=i\delta(x-x')$.

The scaling dimension $x_{n}$ of an operator $O_n$ is related with the energy eigenvalue $E_{n}(N,M^{z})$ of the state generated by applying the operator to the ground state\cite{ca,HW,IA} as, $x_{n}=\displaystyle \lim_{N \rightarrow \infty}x_{n}(N)$, where
\begin{equation}
\label{eq:dim}
x_{n}(N)=\frac{2N}{2\pi v_{\sbox{s}}}[E_{n}(N,M^{z})-E_{\sbox{g}}(N,M^{z})].
\end{equation}
We can identify the correspondence between the operators in boson representation and the eigenstates of the spin chains by comparing their symmetry\cite{nomura,at}. In this way, from the excitation energies of the eigenstates corresponding to the operators  $O_1=\cos \sqrt{2}\phi$ and $O_2=\sin \sqrt{2}\phi$, we can determine their scaling dimensions $x_1$ and $x_2$ using the relation (\ref{eq:dim}). On the other hand, both $x_1$ and $x_2$ should be equal to $K/2$ as determined from their correlation function. Thus the value of $K$ is determined from the numerically obtained values of  $x_1$ and/or $x_2$. Actually, to reduce the finite size correction to $O(1/N^2)$, it is more convenient to use the combination,
\begin{equation}
K(N)=x_{1}(N)+x_{2}(N),
\end{equation}
as proposed by Kitazawa and Nomura\cite{at}. We assume the formula
\begin{equation}
K(N)=K+\frac{c_{1}}{(2N)^{2}}+\frac{c_{2}}{(2N)^{4}}
\end{equation}
to extrapolate $K(N)$ to $N \rightarrow \infty$. The extrapolation procedure is shown in the Fig. \ref{fig5} which gives $K=1.5227 \pm 0.0003$ for $j=0.5$. The error is estimated from the difference between the values extrapolated from the data for $4N=12,16,20,24$ and $4N=16,20,24,28$.

From the symmetry consideration, the effect of quadrumerization can be represented by the term proportional to $\cos \sqrt{2}\phi$ which explicitly breaks the underlying $U(1)$ symmetry. Such a term can be also explicitly derived for $m^z=1/4$ by the usual Abelian bosonization scheme\cite{ft,ca2,mo,to,to2}. Thus the low energy properties of our model can be described by the sine-Gordon model given by,
\begin{eqnarray}
\label{eq:sg}
{\cal H_{\sbox{SG}}}&=& \frac{1}{2\pi} \int {dx \Big[ v_{\sbox{s}}K (\pi\Pi)^{2}+\frac{v_{\sbox{s}}}{K}(\frac{\partial \phi}{\partial x})^{2} \Big]} \nonumber \\
&+& \frac{y_{1}v_{\sbox{s}}}{2\pi a^{2}} \int{dx \cos \sqrt{2}\phi},
\end{eqnarray}

where the $a$ is the lattice constant. The second term of the Hamiltonian generates the energy gap.

From the renormalization group theory and conformal field theory, the critical exponent of the energy gap $\nu$ produced by the $\cos \sqrt{2}\phi$ term is related to its scaling dimension $x_1$ as
\begin{equation}
\nu=\frac{1}{2-x_1}=\frac{1}{2-\frac{K}{2}}.
\label{eq1}
\end{equation}
Figure \ref{fig6} shows the variation of the critical exponent $\nu$ with $j$. The error bars due to the extrapolation procedure are less than the size of the symbols. For $j > 0$, the critical exponent $\nu$ is close to 0.8 and slightly depends on $j$ with a weak maximum $\nu \simeq 0.81$ at $j=1$, while for $j < 0$, the critical exponent $\nu$ increases with $j$ and appraoches 2 as $j \rightarrow -0$. The positive values of $\nu$ ensure that the point $\delta=0$ is the critical point (unstable fixed point) and the plateau opens for infinitesimal $\delta$. These values are consistent with the correlation function exponents $\eta(=1/K)$ and $\eta^z(=K)$ for $j=-2$ and $0.5$ obtained by Sakai\cite{ts1}. The values for large negative $j$ are also consistent with the values of $\eta$ and $\eta^z$ for the $S=1$ antiferromagnetic Heisenberg chain obtained by Sakai and Takahashi\cite{st1} which give $\nu \cong 1.74 \pm 0.1$. Obviously, $\nu=1$ for $j=0$.  It is remarkable that there is a discontinuity in  $\nu$ at $j=0$.
\subsection{Analytic Results for the Critical Exponent $\nu$}

\subsubsection{Undimerized case : $j = 1$}

For $j=1$ and $\delta=0$, the solution by the Bethe's hypothesis is available\cite{bh,bik}. In this case, the parameter $K$ is determined from the  dressed charge to yield $\nu \simeq 0.81$ for $m^z=1/4$  by Fledderjohann {\it et al.}\cite{fled1,fled2} which is consistent with our numerical results.

\subsubsection{Mapping onto the effective unmagnetized XXZ chain for $j \simeq 0$ and $j >> 1$}

The magnetized state of the present model on the plateau can be mapped onto the unmagnetized state of the effective XXZ chain in the limiting cases $j \simeq 0$ and $j \rightarrow \infty$ using the method proposed by Totsuka\cite{to}.

For  small values of $j$,  the dimers are formed on the $1+\delta$-bonds and $1-\delta$-bonds in the ground state, if the magnetic field is absent. When the magnetic field is switched on, some of the dimerized bonds are broken and change into the $\mid \uparrow \uparrow>$ state. Other states on the $1 \pm \delta$-bonds are energetically unfavourable and can be discarded. Then the present system can be described as a gas of the interacting dimers as follows
\begin{eqnarray}
{\cal H}' &=& \frac{j}{4}\sum_{i=1}^{2N} [n_{i}n_{i+1}-(d_{i}^{+}d_{i+1}+d_{i+1}^{+}d_{i})] \nonumber \\
&+& \sum_{i=1}^{2N} (H-1)  n_{i}+\delta \sum_{i=1}^{2N}(-1)^{i} n_{i},
\end{eqnarray}
where $d_{i}^{+}$ and $d_{i}$ are the fermion operators and $n_{i}=d_{i}^{+}d_{i}$. The state with $n_{i}$=1 corresponds to the state in which the $i$-th $1 \pm \delta$-bond is occupied by a dimer, while $n_{i}$=0 to the $\mid \uparrow \uparrow>$ state. This Hamiltonian can be further mapped onto the $S=1/2$ XXZ chains by the Jordan-Wigner transformation with the correspondence
\begin{equation}
n_{i}=\hat{S}_{i}^{z}+\frac{1}{2},
\end{equation}
where $\hat{\v{S}}_{i}$ is the effective spin-1/2 operator. Then, the effective XXZ Hamiltonian is given by
\begin{eqnarray}
{\cal H}'_{\sbox{eff}} &=& \frac{\mid j \mid}{2} \sum_{i=1}^{2N} \Big[\hat{S}_{i}^{x}\hat{S}_{i+1}^{x}+\hat{S}_{i}^{y}\hat{S}_{i+1}^{y}+\Delta \hat{S}_{i}^{z}\hat{S}_{i+1}^{z}\Big] \nonumber \\
&+& \sum_{i=1}^{2N}(H-1)\hat{S}_{i}^{z}+ \delta \sum_{i=1}^{2N}(-1)^{i}\hat{S}_{i}^{z},
\end{eqnarray}
where $\Delta=1/2$ for $j > 0$ and $\Delta=-1/2$ for $j < 0$.
 It should be noted that the staggered field term proportional to $\delta$ appears because the energy required to break the dimer bond is different according whether $i$ is even or odd.

The parameter $K$ can be calculated by the Bethe's hypothesis\cite{bik,lp} as
\begin{equation}
K=\frac{\pi}{\pi-\mbox{arccos}\Delta}.
\end{equation}
Therefore we have $K=3/2$ for $j \rightarrow +0$ and  $K=3$ for $j \rightarrow -0$. In the bosonized language, the staggered field term corresponds to $\sin \sqrt{2}\phi$ term\cite{nomura}. Therefore we can again use the formula (\ref{eq1}) to obtain $\nu$ is 4/5 for $j \rightarrow +0$ and $\nu=2$ for $j \rightarrow -0$.

On the other hand, for $j \rightarrow + \infty$, the dimers reside on the $j$-bonds in the ground state. Using the similar argument as the small $j$ case, the effective XXZ Hamiltonian can be written as
\begin{eqnarray}
{\cal H}''_{\sbox{eff}} &=& \sum_{i=1}^{2N} \{ \frac{1+(-1)^{i}\delta}{2}\}(\hat{S}_{i}^{x}\hat{S}_{i+1}^{x}+\hat{S}_{i}^{y}\hat{S}_{i+1}^{y}+\Delta \hat{S}_{i}^{z}\hat{S}_{i+1}^{z}) \nonumber \\
&+& \sum_{i=1}^{2N} (H-j)\hat{S}_{i}^{z},
\end{eqnarray}
with $\Delta=1/2$ resulting in $K=3/2$. In this case, the dimerization term appears in the effective Hamiltonian which has the bosonized form $\cos \sqrt{2}\phi$\cite{nomura}. This term also has the same scaling dimension $K/2$. Therefore we have again $\nu =4/5$ for $j \rightarrow \infty$.
Thus the analytic results explain our numerical results quite well.
\section{Summary and Dicussion}
The magnetization plateaus at $m^{z}=1/4$ in the one dimensional $S=1/2$ Heisenberg model with dimerization and quadrumerization are investigated by the exact diagonalization of the finite size systems. The central charge at the critical point $\delta=0$ is calculated to be close to unity. Under the assumption of the universality class of the $c=1$ Gaussian model, thus, the critical exponent of the plateau width $\nu$ is obtained by the level spectroscopy method from the numerically calculated energy spectrum. The critical exponent $\nu$ is almost constant and close to 0.8 for $j > 0$, while for $j < 0$, it strongly depends on $j$ and increases up to 2 as $j$ approaches $-0$ as shown in Fig. \ref{fig6}.

For $j\rightarrow \pm 0$ and $j\rightarrow +\infty$, the present model is mapped onto the effective unmagnetized XXZ model based on the dimer gas model. This gives the results consistent with the numerically obtained values of $\nu$. At $j=1$, the result is also consistent with the analytical calculation using the Bethe's hypothesis.

According to the present calculation, the magnetization plateau at $m^z=1/4$ should be always observed in the presence of quadrumerization. Therefore it must be interesting to synthesize the linear chain material which has the present structure. It should be also remarked that it is easier to observe the plateau for positive $j$ rather than negative $j$ as far as the strength of the quadrumerization is weak.

The value of $\nu$ is also related to the exponent of the energy gain $\Delta E$ due to quadrumerization at small $\delta$ as $\Delta E \propto \delta^{2\nu}$. For $j > 0$, $2\nu \sim 1.6 < 2$. Therefore the energy gain due to quadrumerization is larger than the lattice deformation energy and quadrumerization takes place spontanuously at $m^z=1/4$ in the similar way as the spin-Peierls instability. Actually, such instability can take place for arbitrary values of magnetization as far as $\nu <1$. In the case of quadrumerization, the wave number of the spatial modulation is commensurate with the original lattice. Therefore the lattice distortion is pinned to the lattice and the plateau will be stabilized. Observation of such field induced spin-Peierls transition would be also an interesting problem from the experimental side. On the other hand, for $j < 0$, $\nu$ is always larger than unity and no spontanuous quadrumerization is expected.

We are grateful to H. Nishimori for the program package TITPACK version 2 for the diagonalization of spin-1/2 systems. Chen and Hida thank S. Yamaguchi for valuable discussion on the conformal field theory. The numerical calculation is performed using the HITAC S820 and SR2201 at the Information Processing Center of Saitama University and the FACOM VPP500 at the Supercomputer Center of Institute for Solid State Physics, University of Tokyo.

\newpage
\begin{figure}
\caption{Schematic magnetization curve of the present model.}
\label{fig1}
\end{figure}

\begin{figure}
\caption{The critical fileds for (a)$j=5$, (b)$j=0.5$, (c)$j=-0.5$ and (d)$j=-5$. The critical fields $\Hca$, $\Hcb$ and $\Hcc$ are represented by $\bullet$, $\diamond$ and $\circ$, respectively. The solid thick line is the saturation field $\Hs$. The thin solid lines are just guides for eye.}
\label{fig2}
\end{figure}
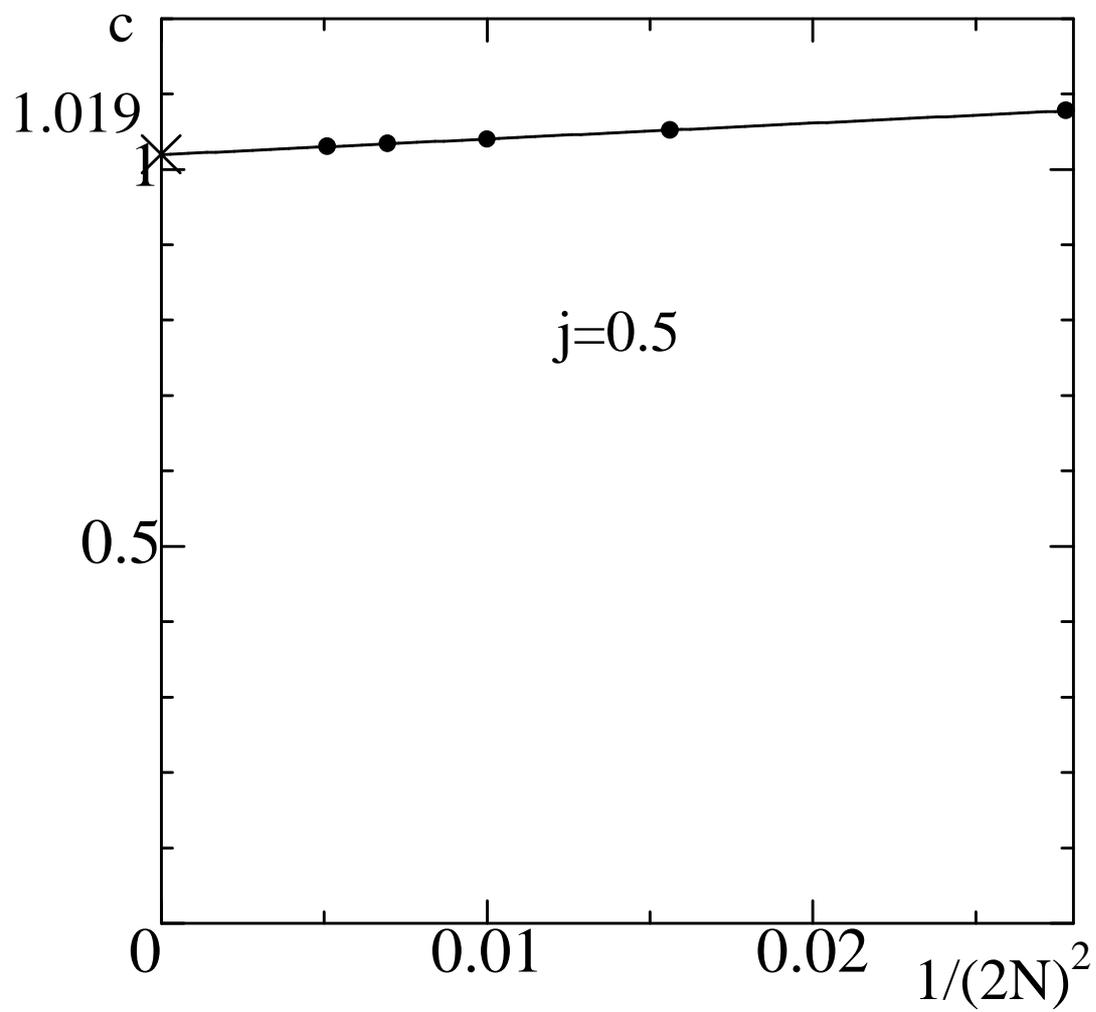
\begin{figure}
\caption{The extrapolation procedure of the finite size central charge $c$ for $j=0.5$.}
\label{fig3}
\end{figure}
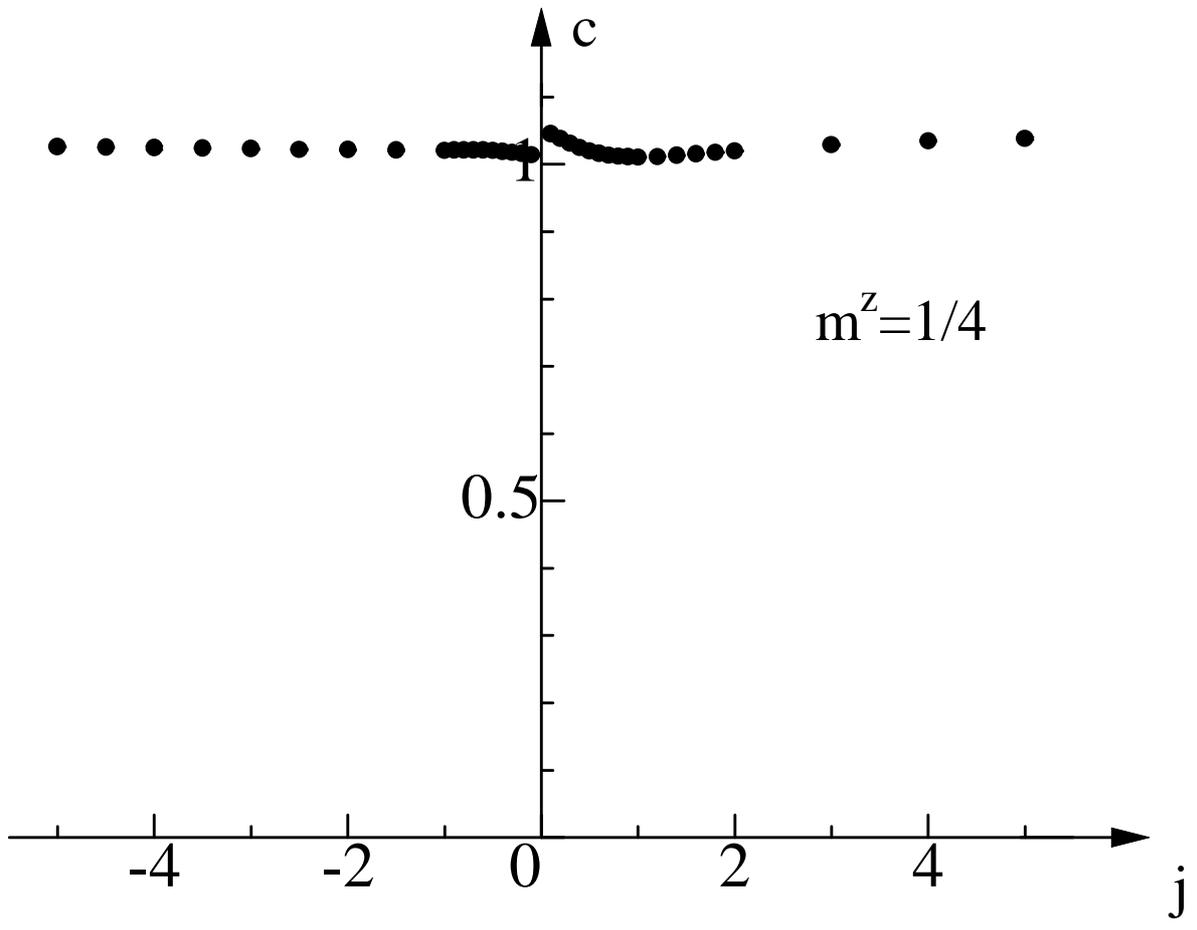
\begin{figure}
\caption{The $j$ dependence of the numerically obtained central charge $c$.}
\label{fig4}
\end{figure}
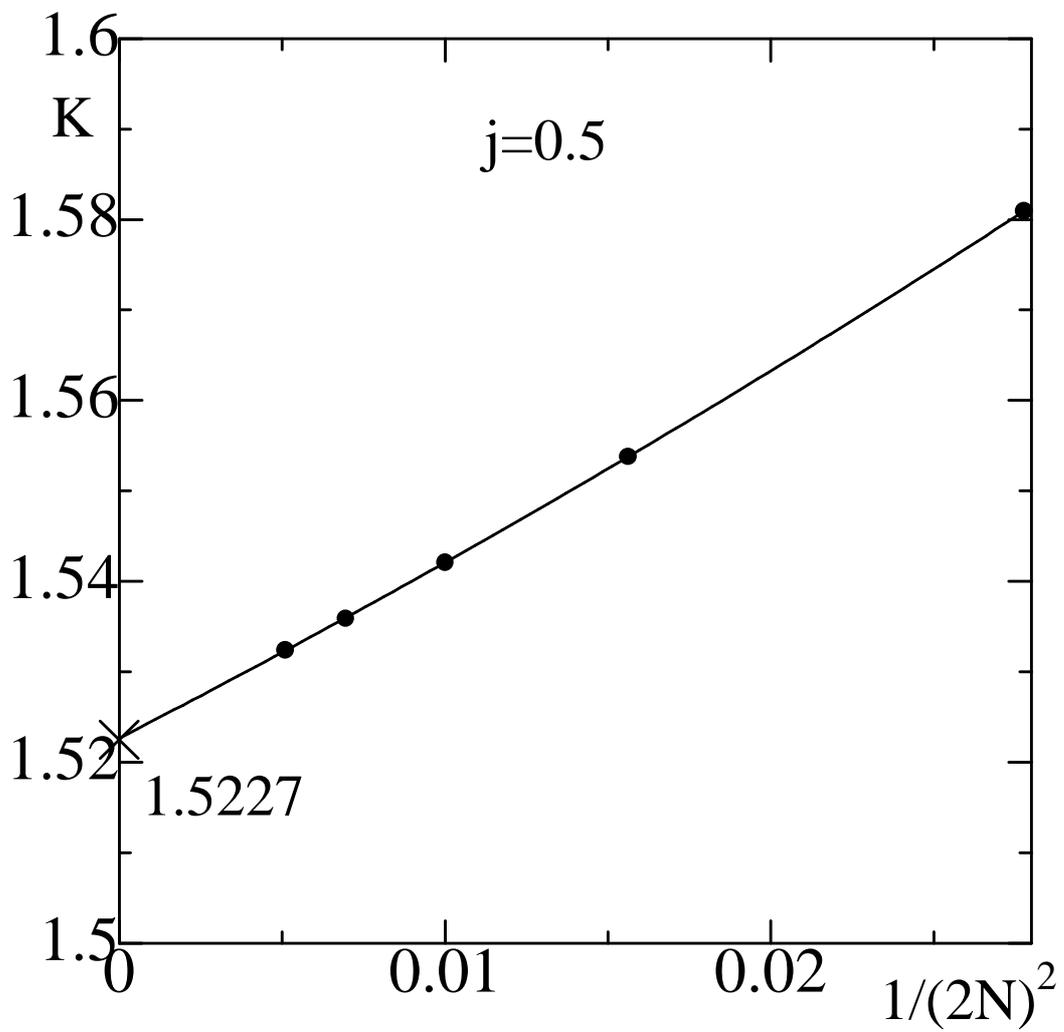
\begin{figure}
\caption{The extrapolation procedure of $K$ at $j=0.5$.}
\label{fig5}
\end{figure}
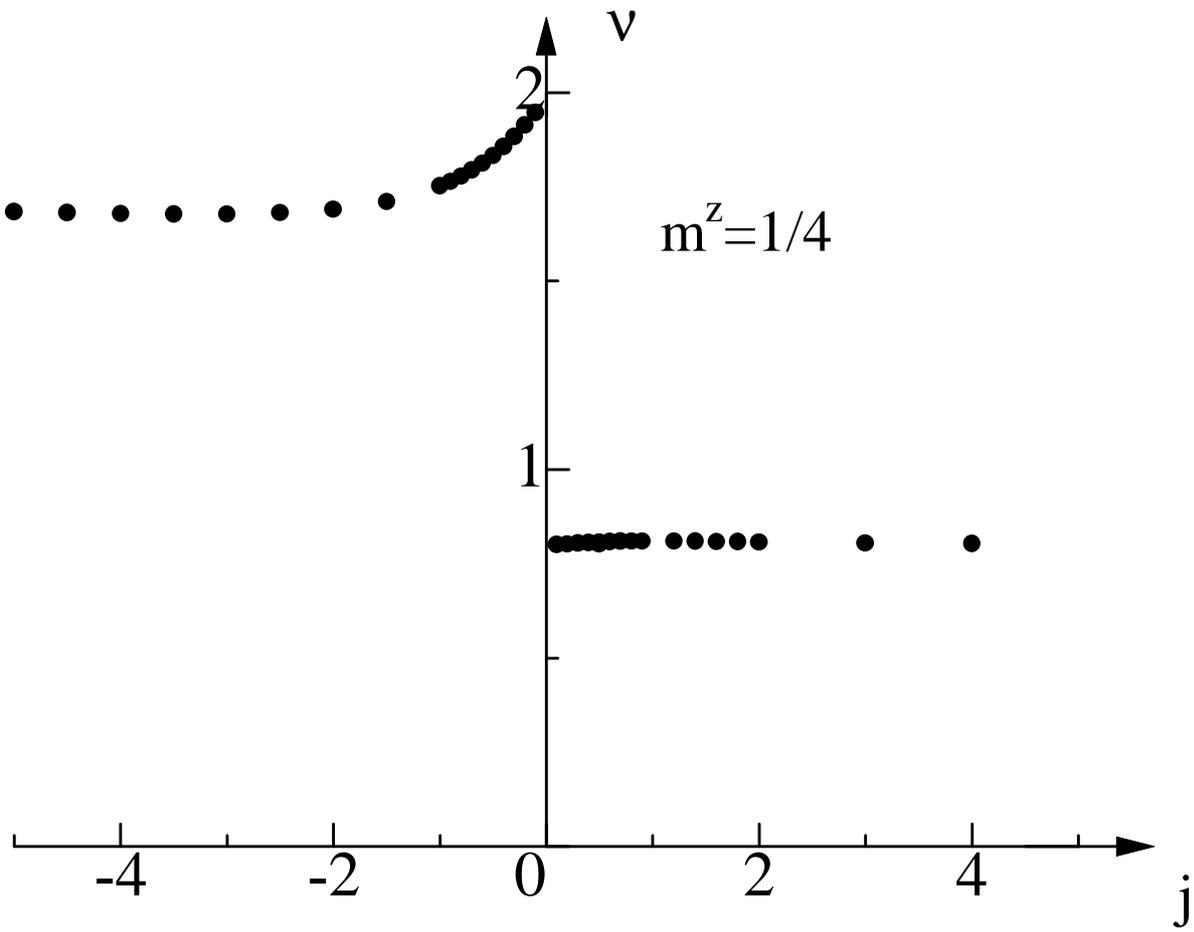
\begin{figure}
\caption{The $j$ dependence of the critical exponent $\nu$.}
\label{fig6}
\end{figure}

\end{document}